\documentclass[reprint,twocolumn,secnumarabic,superscriptaddress,nobibnotes,aps,prr,floatfix]{revtex4-2}

\usepackage[utf8]{inputenc}
\usepackage{graphicx}
\usepackage{tikz}
\usepackage{amsmath,amsfonts,amssymb} 
\usepackage{xcolor}
\usepackage[caption=false]{subfig} 
\usepackage{ragged2e}              
\usepackage[colorlinks=true,citecolor=blue,linkcolor=magenta]{hyperref} 

\definecolor{customblue}{HTML}{219DFF}

\begin{document}

\title{Aging in a two-dimensional swarmalator crystal with delayed interactions}%

\author{Chanin Kumpeerakij}
\email[]{chanin.kumpeerakij@colorado.edu}
\affiliation{University of Colorado, Boulder, CO}
\affiliation{Chula Intelligent and Complex Systems Lab, Department of Physics, Faculty of Science, Chulalongkorn University, Bangkok, Thailand}
\author{Thiparat Chotibut}
\email[Correspondence to: ]{thiparatc@gmail.com}
\affiliation{Chula Intelligent and Complex Systems Lab, Department of Physics, Faculty of Science, Chulalongkorn University, Bangkok, Thailand}
\author{Oleg Kogan}
\email[Correspondence to: ]{oleg.kogan@qc.cuny.edu}
\affiliation{Queens College of CUNY, Flushing, NY}

\begin{abstract}
Time delay can have a significant impact on the properties of collective organization of active matter.  
In the previous paper \cite{PreviousPaper}, we discussed delay-induced breathing in a system of swarmalators -  a model coupling particles’ internal phases to spatial interactions.  Here we build on that study to  
investigate the aging phenomenon in this system.  It is the aging of a two-dimensional crystal with defects and inhomogeneous lattice constants, and takes place after the breathing transients subside.  We show that aging proceeds through the gradual elimination of five-fold and seven-fold coordination number defects, which merge pairwise or migrate to the cluster boundary, incrementally increasing the hexatic order parameter in the bulk.  Despite this process, defects usually do not fully disappear; some residual number of defects remain frozen in the interior.  However, we found that it is possible to achieve a nearly total elimination of coordination number defects at sufficiently low delay - when the surface of the cluster develops a sufficeintly thick fluidized ``boiling layer''.  This mechanism of boundary-mediated annealing reveals a non-equilibrium pathway to achieving high crystalline order in the bulk, and raises a tantalizing possibility for controlling defects in active matter with free boundaries.

\begin{description}
\item[Keywords]
swarmalators, swarming, synchronization, topological defects, active defects,  \\ active matter, delayed interactions, non-reciprocal interactions, aging, 2D crystallization 
\end{description}
\end{abstract}

\maketitle

\section{Introduction}
\label{sec:intro}
Active matter offers a rich avenue for investigating collective phenomena far from equilibrium. From bacterial colonies to synthetic colloidal suspensions, active particles are continually driven out of equilibrium, and can often self-organize into new types of spatio-temporal collective patterns~\cite{Ramaswamy2010,Marchetti2013,Vicsek2012, Gompper2020, Shaebani2020, Kevin2017, bowick2022symmetry, shankar2022topological}.  The Landau-Peierls argument, closely related to the Mermin-Wagner theorem for continuous symmetries, predicts that thermal fluctuations destroy true long-range positional order, allowing at best quasi-long-range order~\cite{landau_lifshitz_stat_phys}. However, in active matter, persistent propulsion, delayed feedback, and nonreciprocal interactions can relax these equilibrium constraints, giving rise to a broader repertoire of collective states, such as active turbulence, hyperuniform order, long-range order in 2D, and nontrivial defect dynamics~\cite{Fruchart_Nature2021, Doostmohammadi2016, Wensink_PNAS2012, Irvine_PRX2018, Weijs_PNAS2015, toner1998flocks}.  Investigating how 2D crystallization proceeds under these nonequilibrium conditions remains an active area of research in condensed matter physics.

Within this broader context, the swarmalator model introduced by O’Keeffe, Hong, and Strogatz~\cite{Kevin2017} provides a minimal setting to investigate how active particle positions and internal phase dynamics can coevolve to form out-of-equilibrium collective states.  Swarmalators couple swarming (the spatial attraction or repulsion of agents) with dynamics of internal periodic phase-like variables (which allow phase synchronization or anti-synchronization), leading to states where the geometric, spatial arrangement and phase alignment become intertwined. Recently, Ref.~\cite{PreviousPaper} showed that adding a constant time delay to the phase interaction in the O’Keeffe, Hong, and Strogatz model can produce collective behaviors completely different from those found in the delay-free system. 

Time delays capture realistic features of biological systems, where internal signaling cannot be assumed to act instantaneously.  One scenario that could lead to delay is the processing of received signals and production of signaling molecules by cells \cite{Petrungaro}.  Another scenario that could plausibly lead to time delay is the chemical signaling of internal cell cycle (phase) between cells - if this information propagates slowly on the time scale of cells' internal phase dynamics.  This ``light cone'' type of delay will depend on the relative positions of cells.  In this paper, we consider a simple case of a constant delay, regardless of relative positions - so it is more fitting to the first scenario.

In this system, sufficiently large delays give rise to a pronounced breathing transient: the entire disk-shaped swarm cluster expands and contracts radially, reminiscent of a breathing mode (see Fig.~1 of \cite{PreviousPaper}), characterized by a well-defined oscillation frequency and decay rate. After this breathing subsides, depending on the delay time, the cluster may settle into either what seems like a quiescent state or a dynamic state with persistent boiling-like motions of particles near the boundary. More subtly, even seemingly quiescent clusters exhibit aging dynamics - a slow and long-lived process wherein residual creeping motions gradually nudge particles' positions slightly on ever-longer timescales. Such aging is characterized by the increasingly slow, progressive elimination of local coordination-number defects in the swarmalators’ arrangement.  Thus, aging in swarmalator clusters add to the list of already known glassy relaxation processes observed in other active or soft-matter systems~\cite{Berthier2013,paul2023dynamical, ciletti2003universal, janssen2020aging, flenner2019glassy}.

In this work, we study in detail such late-time aging dynamics numerically, revealing the processes by which the fraction of coordination number defects decreases over time.  Local changes in coordination numbers, particularly fivefold and sevenfold defects, take place through annihilation, migration, or pair-production. Our analysis also shows that when the boiling layer is sufficiently thick, it is possible for the solid core located underneath this layer to be nearly or even completely defect free.  

In passive 2D systems, crystalline melting proceeds via dislocation- and disclination-unbinding described by Kosterlitz-Thouless-Halperin-Nelson-Young (KTHNY) theory \cite{Kosterlitz_1972,Kosterlitz_1973,HalperinNelson_PRL78,NelsonHaperin_PRB79,Young_PRB79}. Our active cluster does not melt; it ages toward higher order, so we invoke KTHNY only as a reference language for defect phenomenology. Our system is non-Hamiltonian and has free boundaries, which act as a sink for defects.  These distinctions place its aging pathway outside the KTHNY framework, akin to the non-equilibrium ordering reported in active Brownian disks and other driven 2-D systems ~\cite{Cugliandolo_2Dstuff, Irvine_PRX2018, Cates_PRL2023}.

The remainder of the paper is organized as follows. In Sec.~\ref{sec:delay_review}, we briefly review the delayed swarmalator model and its previously observed long-lived breathing modes. Section~\ref{sec:aging} focuses on the slow creep and glassy relaxation that persists long after the breathing has decayed. We show that defect annihilation and migration underlie this aging process, leading to gradually improved hexatic order. In Sec.~\ref{sec:annealing}, we study the dependence of the fraction of defects on the quality factor and also discover the possibility of a complete elimination of defects from a solid core underneath the boiling layer. Finally, we conclude in Section \ref{sec:discussion} by situating these findings in the larger landscape of non-equilibrium active matter and discussing open directions for future research.

\section{Swarmalators with delayed interactions}
\label{sec:delay_review}
Consider a system of particles in two dimensions with positions $\mathbf{x}$ and internal phases $\theta$ that evolve in time according to 
\footnotesize
\begin{eqnarray}
\label{eq:spatial} \dot{\mathbf{x}}_i &=&   \frac{1}{N} \sum_{ j \neq i}^N \Bigg[ \frac{\mathbf{x}_j - \mathbf{x}_i}{|\mathbf{x}_j - \mathbf{x}_i|} \Big( 1 + J \cos(\theta_j(t-\tau) - \theta_i(t))  \Big)  -   \frac{\mathbf{x}_j - \mathbf{x}_i}{ | \mathbf{x}_j - \mathbf{x}_i|^2}\Bigg]   \label{eq:position_eq} \nonumber \\ \\
\label{eq:phase} \dot{\theta_i} &=&  \frac{K}{N} \sum_{j \neq i}^N \frac{ \sin(\theta_j(t-\tau) - \theta_i(t))}{ |\mathbf{x}_j - \mathbf{x}_i| }. \label{eq:theta_eq}
\end{eqnarray}
\normalsize
Here $\tau$ is the time delay - a particle $i$ responds to the phase of a particle $j$ as it was time $\tau$ ago.  The parameter $J$, which is always chosen to be positive, determines the degree to which similarity of phases increases the tendency of two particles to move towards each other.  The parameter $K$ is the phase coupling parameter that controls the tendency of particles to align phases.  Positive $K$ causes particles to align their phases, while negative $K$ causes anti-alignment.  Overall, Eqs.~(\ref{eq:spatial})-(\ref{eq:phase}) describe a two-way feedback between the spatial and phase organization: relative phases affects spatial dynamics, and spatial positions affect phase dynamics.   With $\tau$ set to zero, the equations reduce to the O’Keeffe, Hong, and Strogatz model \cite{Kevin2017}.  The role of non-zero $\tau$ was studied in \cite{PreviousPaper}.  Before focusing on the long time aging dynamics, we first review the relevant phenomenology introduced by the time delay.      

\subsection{Transient state: Breathing behavior}
Following an initial condition (see Section III.A of \cite{PreviousPaper}) the particles undergo a complicated transient motion.  During an early stage of the transient the particles experience spatial rearrangements.  With the passage of time, the cluster of particles enters the ``breathing behavior'' in which the radius of each particle oscillates - see Fig.~\ref{fig:Fig1}.  In the later stages of this breathing behavior the particles no longer rearrange, i.e. the identity of neighbors, as defined by Delaunay triangulation, for example, becomes fixed.  After the breathing decays away, the cluster of particles enters either a boiling state or a quasistatic state, depending on the value of $\tau$ \cite{PreviousPaper}.

\subsection{Long time phases}
In the boiling phase, the cluster of particles separates into two regions - the fluidized boiling layer where particles are visibly moving in convective-like motions, and a solid core underneath this fluidized layer, where particles are visibly stationary.  This type of phase is found for a sufficiently low delay, below a certain critical value $\tau_c$ \cite{PreviousPaper}.  On the other hand, for $\tau > \tau_c$, the entire cluster is visibly solid, i.e. there is no boiling layer on the surface.  In both cases, the solid part of the cluster is not fully hexagonal - it contains patches of hexagonal crystalline order, but there are defects that make it an imperfect crystal.   
\begin{figure}[h]
\centering
\subfloat[Average speed as a function of time.  The breathing behavior ends at $t \approx 300$. Inserts show a velocity pattern: left - expansion during one of the ``breaths''; right - during the slow creep (the magnitudes of velocity vectors have been greatly increased for visibility).  The centers of clusters were placed at the origin.]{%
\includegraphics[width=\linewidth]{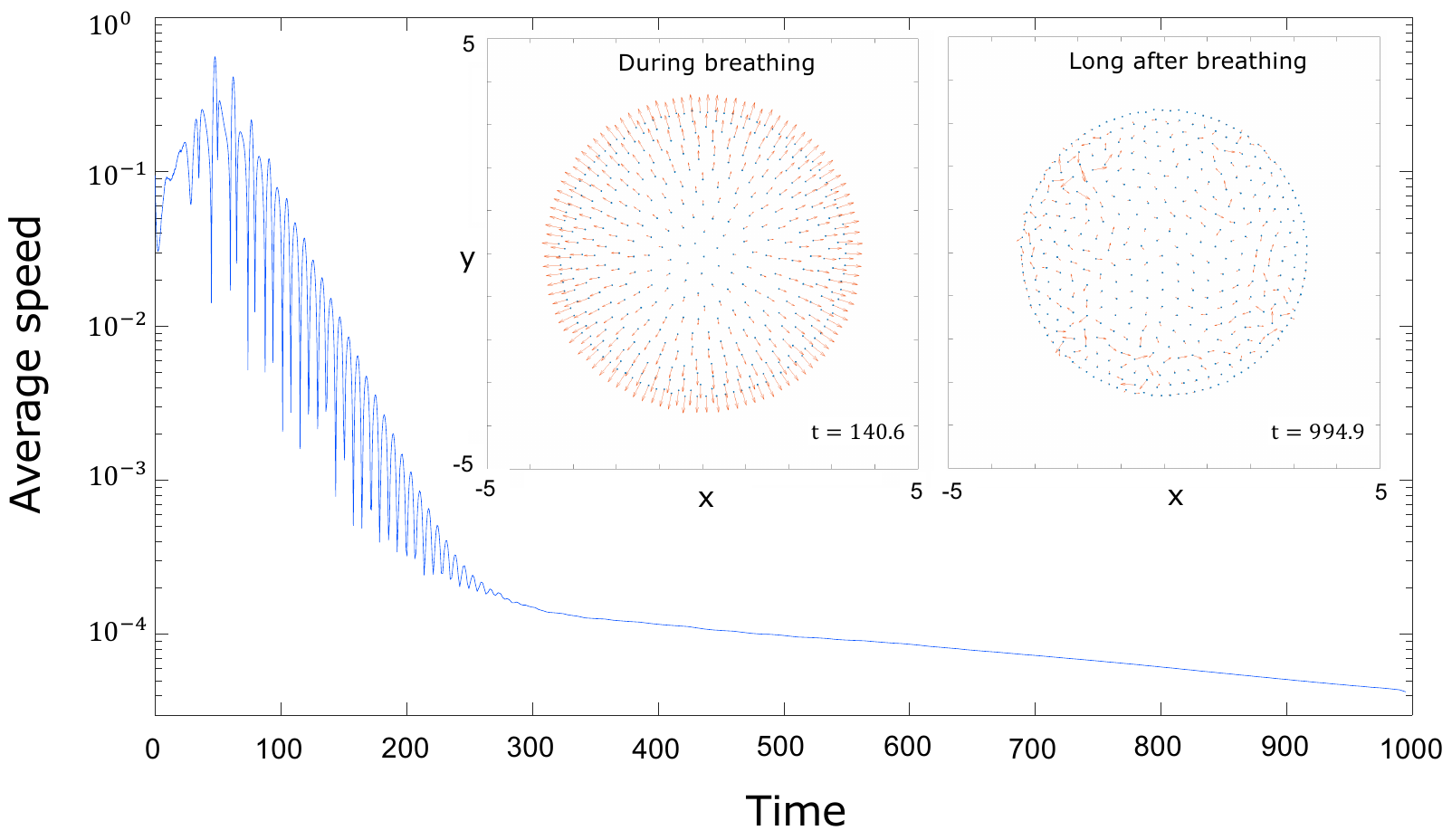}
}\par
\subfloat[Average radius as a function of time during breathing.]{%
\includegraphics[width=\linewidth]{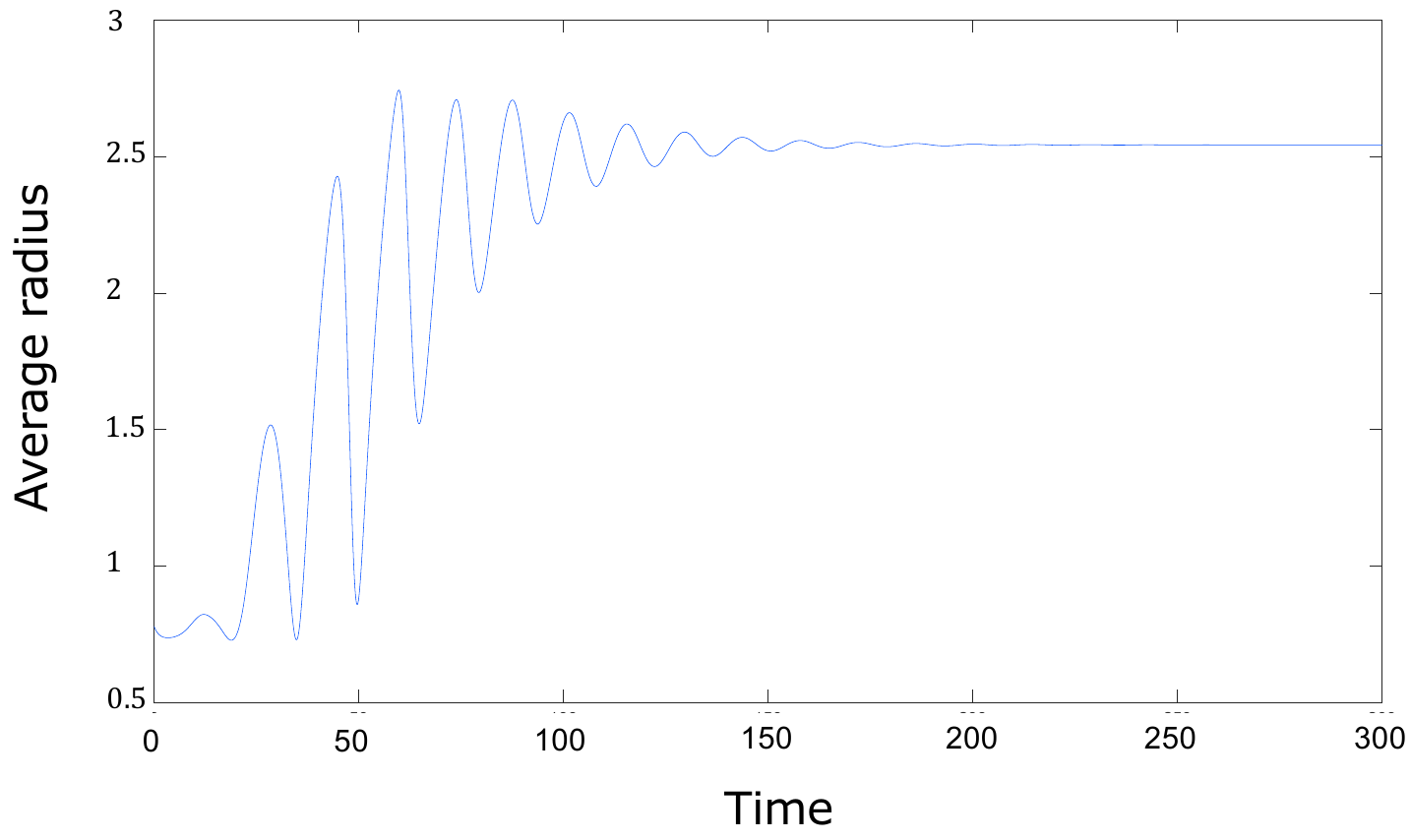}
}
\caption{Breathing transient and the slow creep that follows it - an example.  Here $N=400$, $J=1$, $K=-0.75$, $\tau=11$.}
\label{fig:Fig1}
\end{figure}
When we examine the particle velocities on a logarithmic scale for $\tau > \tau_c$, we realize that there are residual velocities - hence the adjective quasistatic.  These velocities undergo certain dynamics, with an overall tendency to decrease in magnitude, see Fig.~\ref{fig:Fig1}.  We go into a detailed description of this aging process in Section \ref{sec:aging}. 

As $\tau$ is lowered below $\tau_c$, the thickness of the boiling layer increases with decreasing $\tau$.  Concurrently, the properties of the breathing behavior also change with decreasing $\tau$.  Specifically, the frequency $\omega$ and decay rate $\gamma$ of breathing both tend to increase.  The decay rate, however, increases faster, so the quality factor of breathing, defined as $Q \equiv \frac{\omega}{2\gamma}$ decreases, and eventually tends to zero (the quality factor of an oscillator is a measure of the number of oscillations during a decay time, i.e. it is a characteristic number of oscillations).  The quality factor goes to zero at a lower critical delay, which we call $\tau_l$. It appears that the thickness of the boiling layer becomes the entire radius of the cluster around the same $\tau$, and the phases stop synchronizing also around the same $\tau$.  

\begin{figure}[h]
\centering
\subfloat[Angular frequency $\omega$ (blue diamonds) and decay rate $\gamma$ (red triangles) as functions of delay. ]{%
\includegraphics[width=\linewidth]{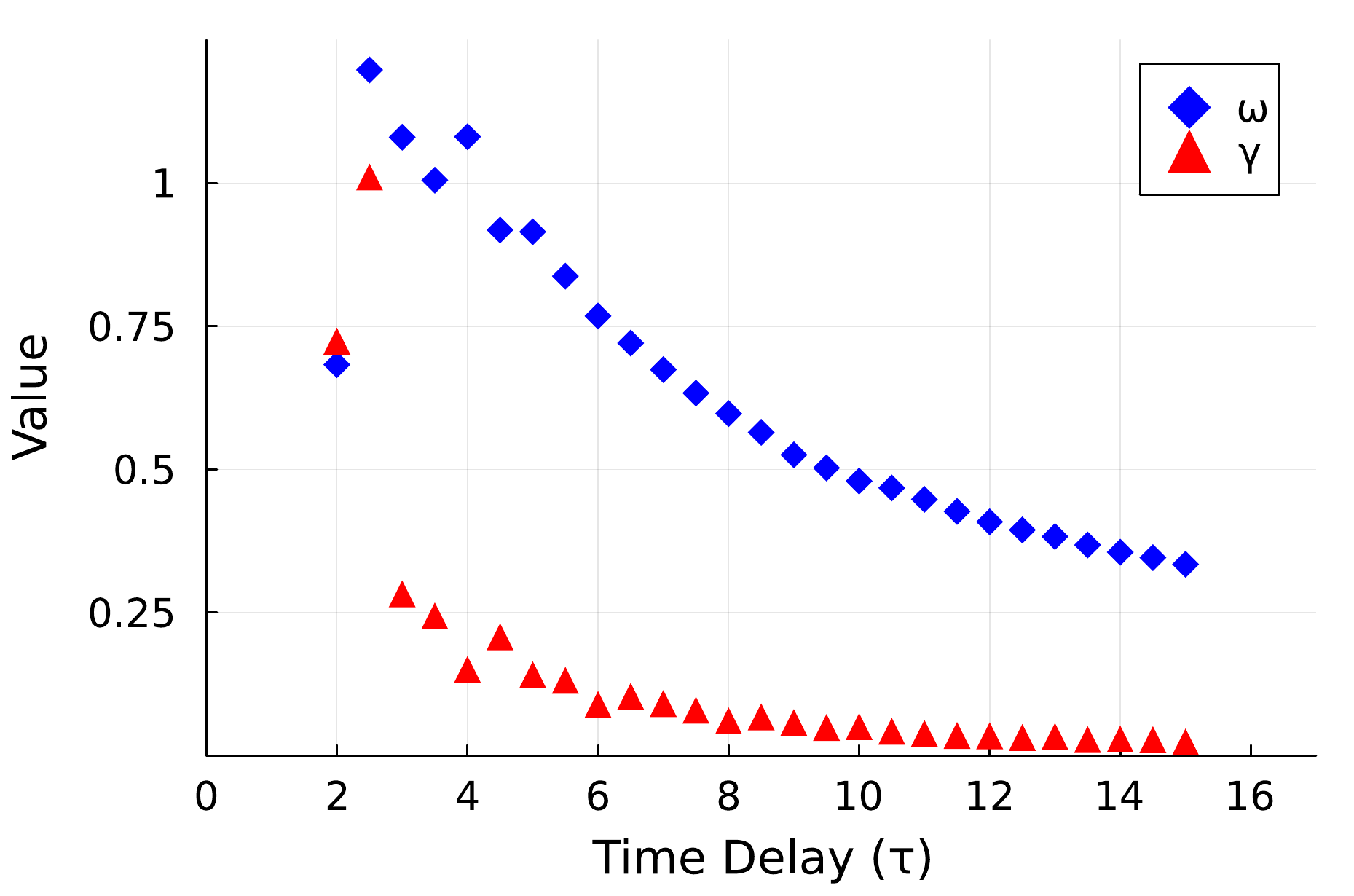}
\label{fig:omega_lambda}}
\par
\subfloat[Quality factor $Q\equiv \frac{\omega}{2\gamma}$ as a function of delay.]{%
\includegraphics[width=\linewidth]{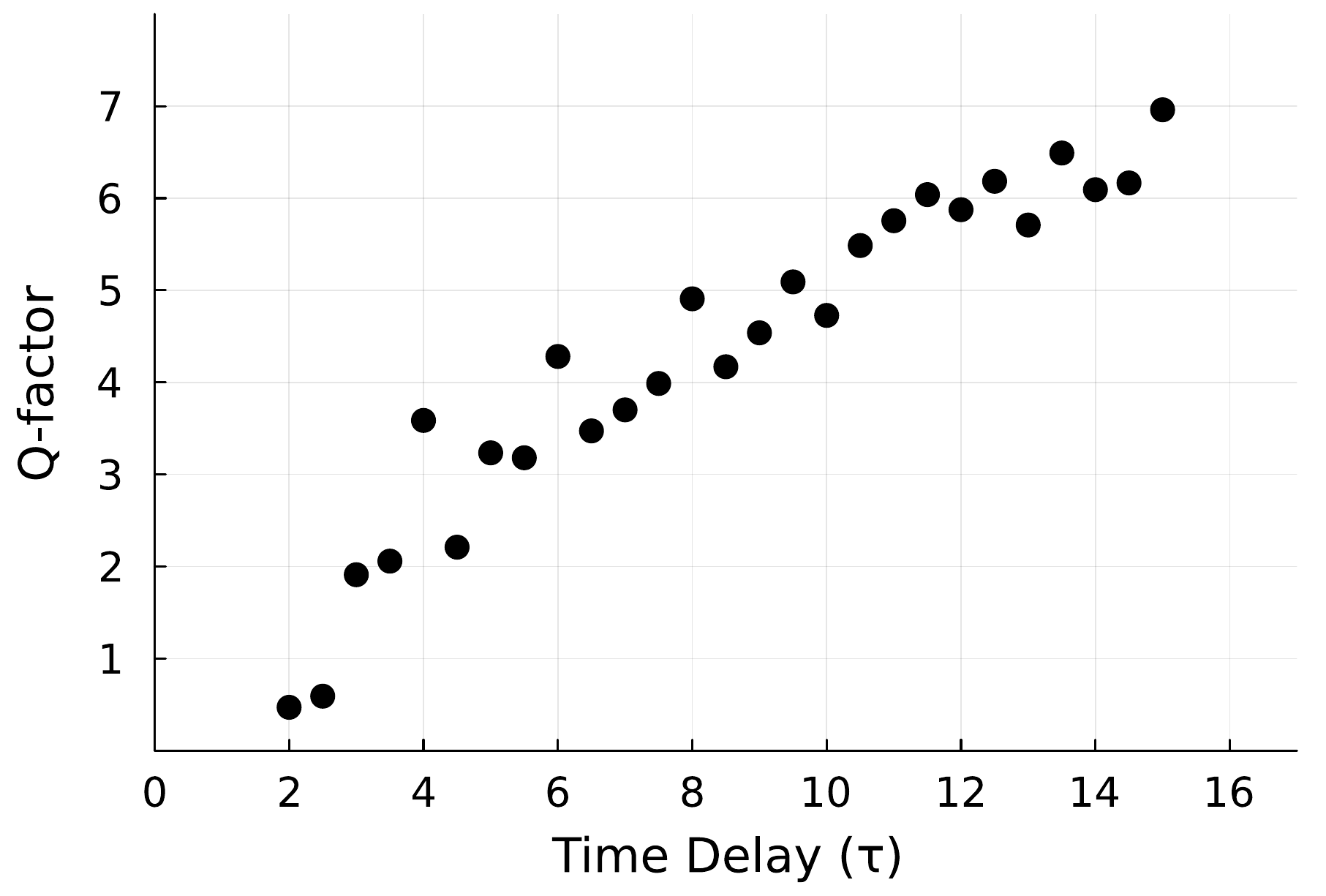}
\label{fig:q-factor_tau}}
\caption{Properties of breathing oscillations.  In this example, the parameters are $N=1000$, $J=1.0$, $K=-0.75$. The lower critical $\tau$ is evidently close to $2.0$.}
\label{fig:Fig2}
\end{figure}
We have not studied the details of the latter two processes to identify whether they happen at exactly the same $\tau$, and this will not be addressed in the current paper.

\section{Aging dynamics}
\label{sec:aging}
\begin{figure}[h]
\centering
\includegraphics[width=0.5\textwidth]{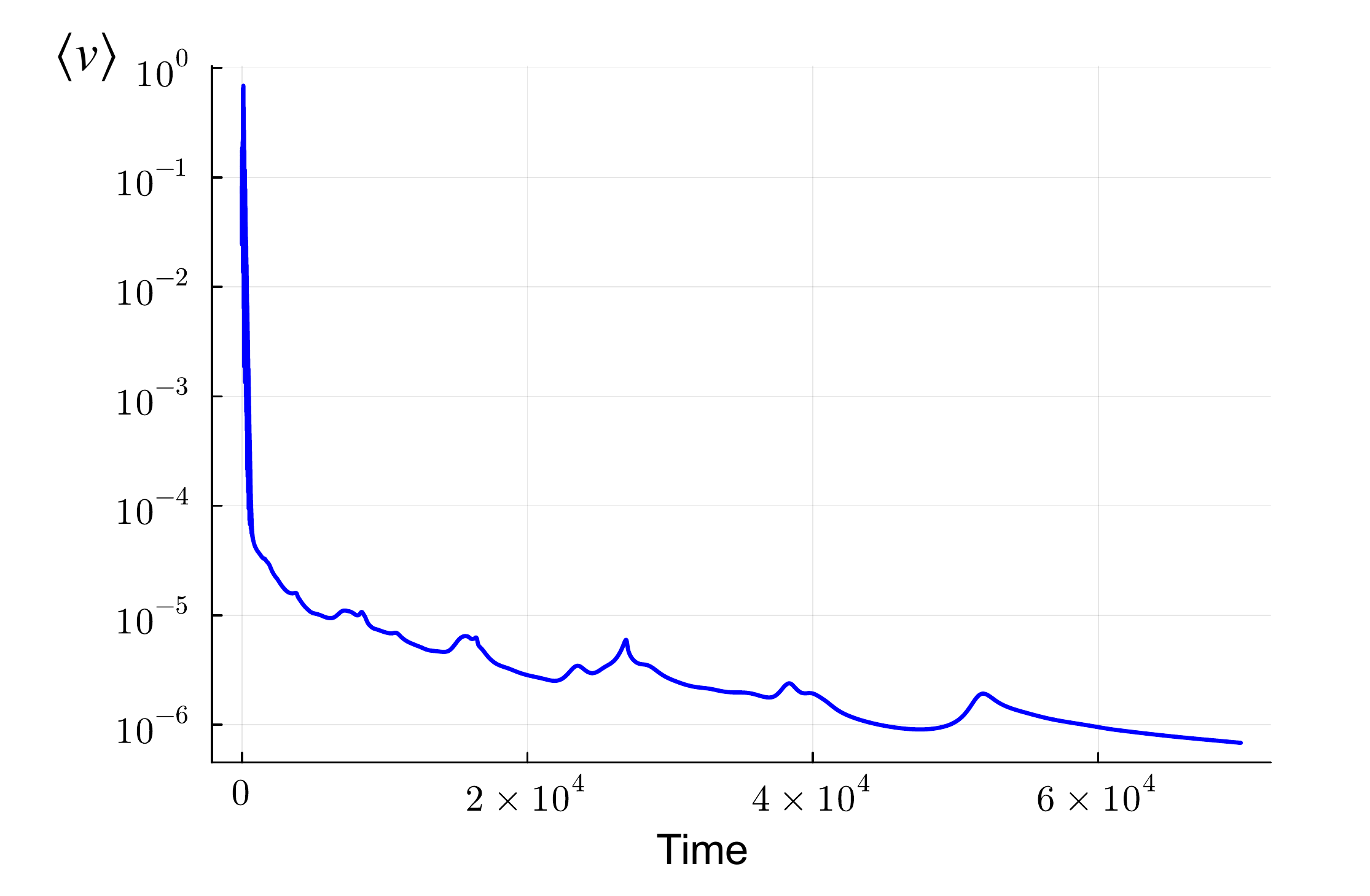}
\caption{Average speed over all particles as a function of time. Here the time span is two orders of magnitude longer than the time to complete the breathing (the breathing regime is seen as a vertical drop at the very beginning of the graph; the oscillations are not visible because they finish at time much less than the total time span shown).  This plot demonstrates that the relaxation is not uniform when observed over such a long time span, but consists of occasional bursts of increase in the average speed. The parameters are $N=1000$, $J=1.0$, $K=-0.75$, $\tau=20.0$.}
\label{fig:avgV_st}
\end{figure}

We now turn our attention to the aging dynamics, which unfolds over timescales much longer than the decay time of the breathing transient.  Simulation details pertinent to this and following sections can be found in Appendix \ref{sec:SimDetails}. Aging phenomena, characterized by exceedingly slow structural rearrangements, have been documented across a variety of active matter systems~\cite{janssen2020aging, paul2023dynamical, flenner2019glassy, Paoluzzi2024glassinessactive, ciletti2003universal, janssen2022agingthermal, janssen2024rejuvenation}.  In active systems, slow relaxation toward crystalline order can proceed via defect-driven dynamics that differ fundamentally from the near-equilibrium mechanisms described by KTHNY theory~\cite{Digregorio2018phasediagram, Cates_PRL2023, Singh_PRL_Universal2019, Irvine_PRX2018}. Here, we report a slow, incremental defect-healing process that we found in the swarmalator system.  As illustrated in Fig.~\ref{fig:avgV_st}, the average particle speed gradually decreases in a non-monotonic manner, punctuated by occasional increases.  When we look at the cluster itself and velocity vectors, we discover that these increases correspond to episodes of increased mobility (velocity ``hot spots")  
associated with localized adjustments of particle positions that generally lead to the change in the local coordination numbers. However, these are very small adjustments, and they do not lead to exchanges of particle positions (i.e. particle rearrangements).
\subsection{Hexatic Order Parameter}
We also examined the local hexatic order parameter (OP), which is defined for each particle $j$ as 
\begin{equation}
\label{eq:Psi_def}
\Psi(j) = \frac{1}{N_j} \sum_{k=1}^{N_j} e^{6i\theta_k}
\end{equation}
The sum is performed over $N_j$ nearest neighbors of particle $j$; the angle $\theta_k$ is between the vector from particle $j$ to its $k$th neighbor, and some chosen reference axis (the ``$x$'' axis).  The hexatic order parameter is complex, so $\Psi(j) = |\Psi(j)| e^{i\phi(j)}$.   A particle surrounded by a perfectly hexagonal arrangement of neighbors will have $|\Psi| = 1$, while six neighbors spaced at unequal angles will lower $|\Psi|$;  a particle with coordination number $\neq 6$ will have $|\Psi|$ exactly zero if the angles are equally spaced, and non-zero (but $< 1$) if the angles are non equally spaced.  In the remainder of this text, we will occasionally refer to the magnitude of $\Psi$ as ``hexaticity''.

\begin{figure*}[h!tbp]
\centering
\includegraphics[width=1.0\textwidth, keepaspectratio]{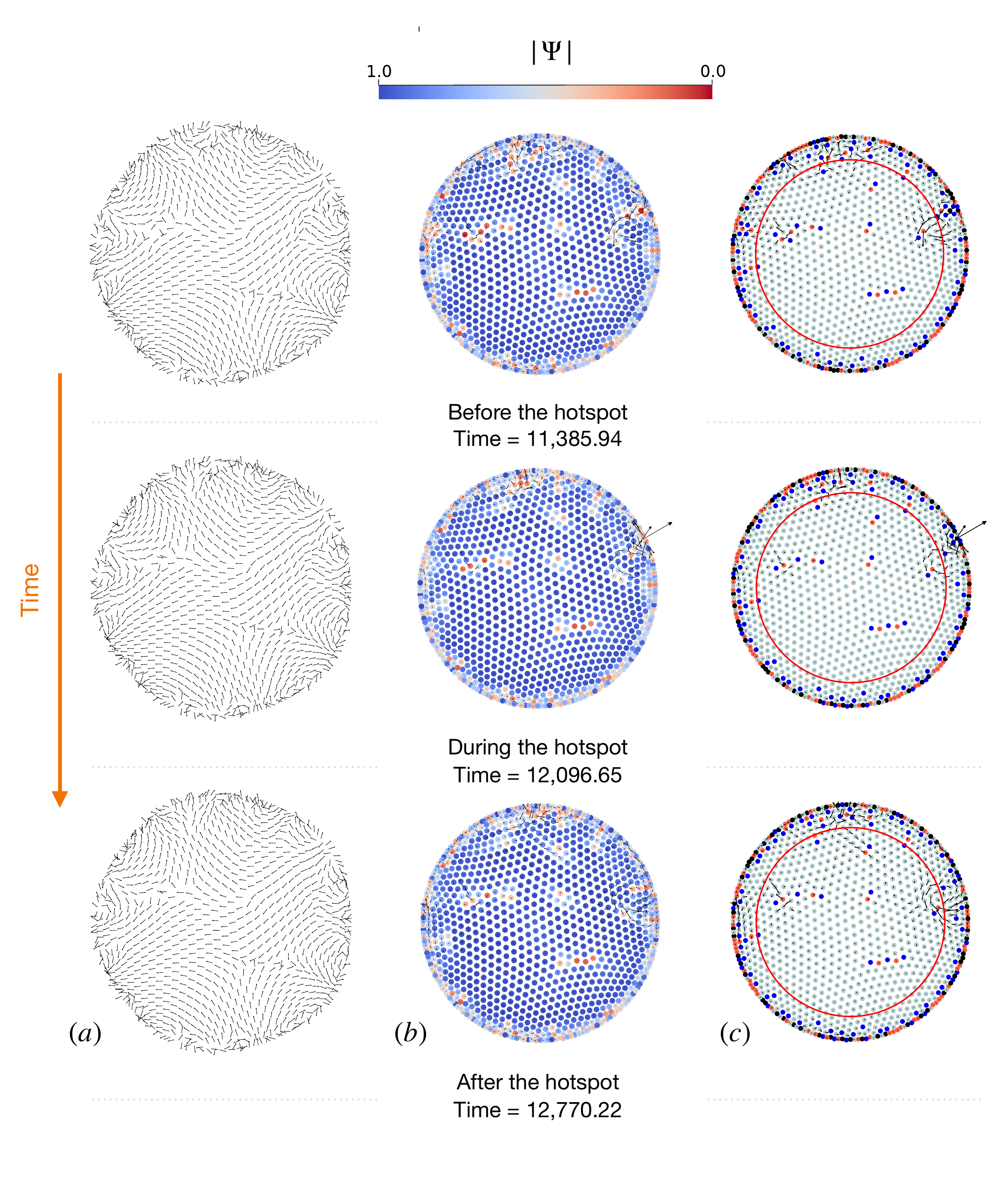}
\caption{An example of a time evolution of a system before, during, and after a velocity hotspot.  (a) Phase of the local hexatic order parameter.  By inspection, we see the presence of topological charges with value $-1$. (b) Velocity hotspots and magnitude of the local hexatic OP.  The color bar represents the scale of absolute magnitude of the local hexatic OP of each particle. (c) Coordination numbers.  Orange particles have 5 nearest neighbors, and blue particles have 7.  Particles with 6 nearest neighbors are in gray, and everything else is black. Here $N=1000$, $K=-0.75$, $J=1.0$, $\tau = 20.0$.  \href{https://o365coloradoedu-my.sharepoint.com/:v:/g/personal/chku8451_colorado_edu/ETMGSMHbAf5ImlFJBXgitzUB9tE3rrig59jxCiwcMLCBZw?e=ucA99i}{[See Supplementary Video 1]}}
\label{fig:hexaticity_evolution}
\end{figure*}

We display in Fig.~\ref{fig:hexaticity_evolution} (b) (middle column) three snapshots of the cluster.  On these snapshots, we add velocity vectors to each particle, and indicate the magnitude $|\Psi_j|$ with color.  The local hexaticity immediately after a velocity hot spot is larger than immediately before it.  The velocity hot-spots ``cure'' the local hexaticity.  Physically, they correspond to microscopic readjustment of positions that lead to increase of the local hexaticity.  We will see below that such positional adjustments can change the identity of neighbors (defined through Delaunay triangulation), but don't involve particle rearrangements.

We visualize the phase part of the local hexatic order parameter in Fig.~\ref{fig:hexaticity_evolution}(a) (left hand column) as arrows that correspond to the argument of $\Psi(j)$, i.e. they depict the complex number $e^{i\phi(j)}$ (we will refer to these as ``phase arrows'').  Consider, for example, the group of three particles with a low $|\Psi|$ in the lower-right portion of the interior of the cluster.  Performing a walk around these three particles in the clockwise direction and completing one revolution, the phase arrows rotate one revolution counter-clockwise.  This is characteristic of topological charges with value $-1$.  The other group of particles with low $|\Psi|$ in the upper-left portion of the interior have the same property of the phase field around it, also characteristic of topological charges with value $-1$.  On the other hand, it is hard to discern the value of the topological charge of defects located at the edges, since the pattern of phase arrows is too complicated there.

\begin{figure}[h]
\centering
\includegraphics[width=1\linewidth]{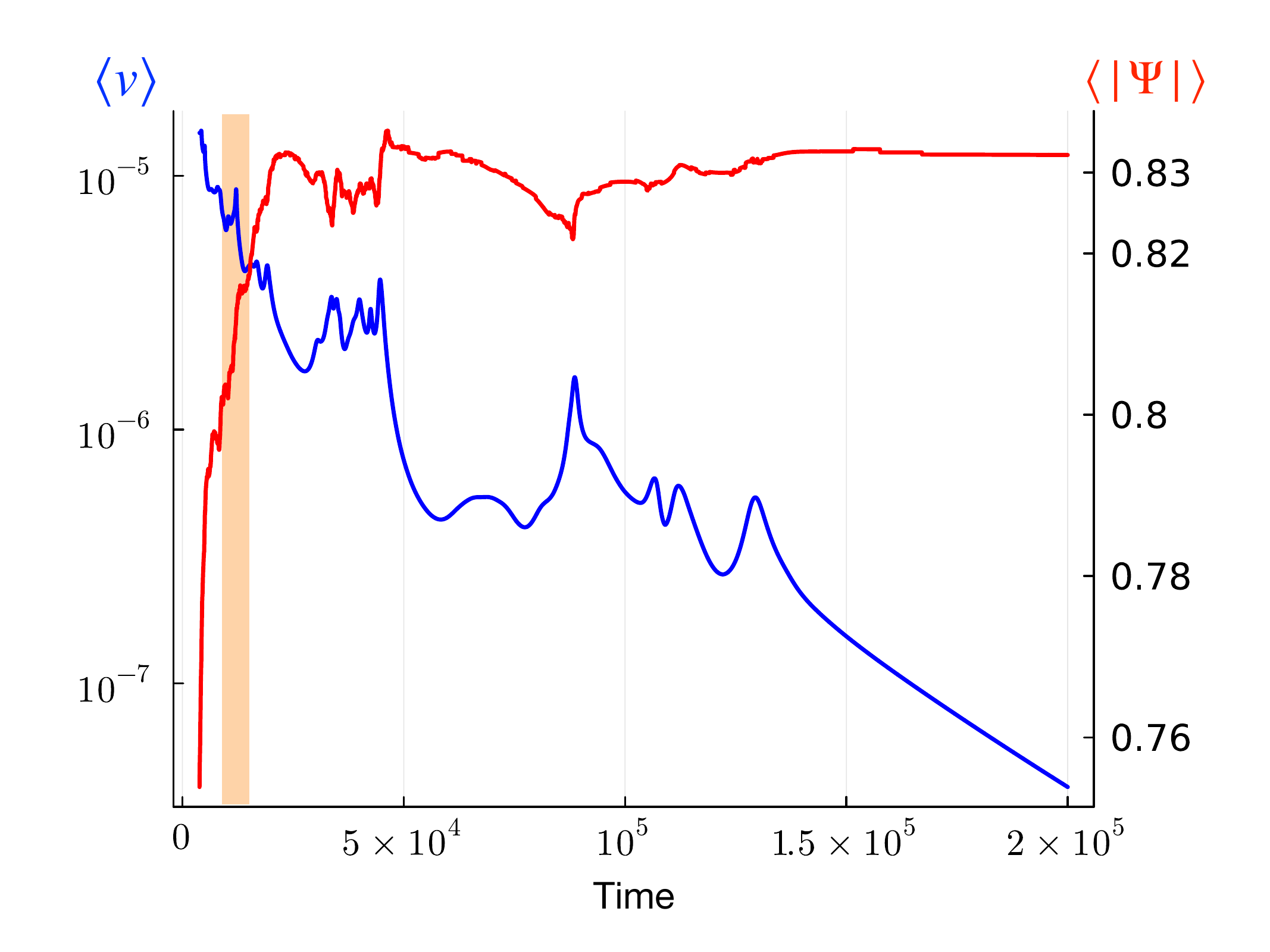} 
\captionsetup{justification=raggedright}
\caption{Evolution of the speed averaged over all particles (blue, decreasing curve), and the local hexaticity averaged over all particles (red, increasing curve).  The highlighted window of time corresponds to Fig.~\ref{fig:hexaticity_evolution}.  Note the uptick in the average speed during that time window. Here $N=1000$, $J=1.0$, $K=-0.75$, and $\tau=20.0$.}
\label{fig:velocity_and_Psi}
\end{figure}

Fig.~\ref{fig:velocity_and_Psi} plots on the same time scale the average velocity and the magnitude of the global hexaticity, which we define as  average of the local $|\Psi(j)|$ over all particles (i.e. it is the average magnitude of the local hexatic order parameter).  We note that while this global hexaticity gradually increases, it does not tend to $1$;  the system does not reach the perfect global hexatic order.  

\subsection{Coordination number defects}
\begin{figure}[h]
\includegraphics[width=1.0\linewidth, keepaspectratio]{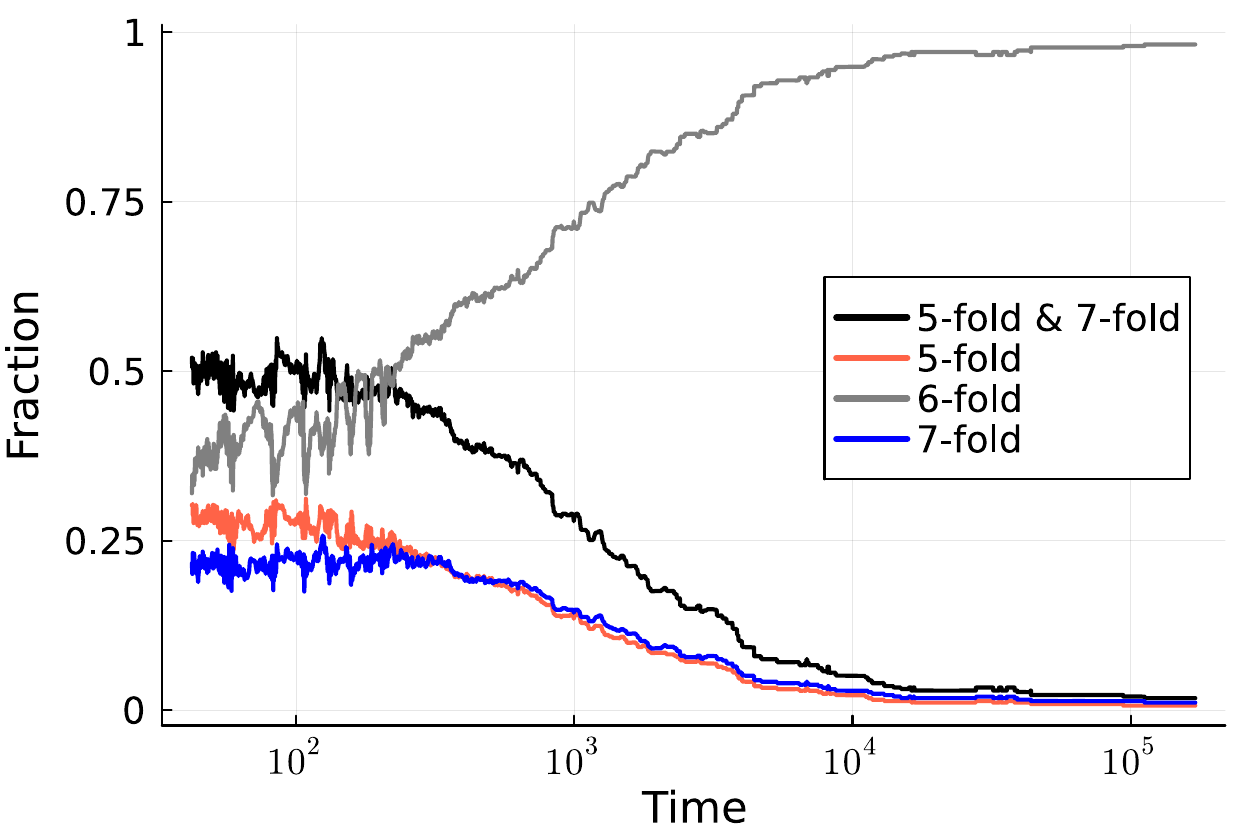}
\caption{Evolution of the fraction of particles that have coordination numbers $5$ (defect, orange), $6$ (black), and $7$ (defect, blue).  The black curve shows the fraction of particles that constitute either $5$- or $7$-defects. Here $N=1000$, $J=1.0$, $K=-0.75$, and $\tau=20.0$.}
\label{fig:cn_evolution}    
\end{figure}

Additional insight into the behavior just described comes from examining local coordination numbers.  To calculate these numbers, we perform Delaunay triangulation, and count the number of nearest neighbors of a given particle.  A particle with the number of nearest neighbors not equal to six is considered a coordination number defect.  It is a more coarse-grained measure than $|\Psi|$ of that particle, since it is possible to have a $|\Psi|<1$ with six neighbors if the neighbors are not spread uniformly, angle-wise.  We give an example of the coordination number plot for in Fig.~\ref{fig:hexaticity_evolution}(c) (right-hand column).  We also display in Fig.~\ref{fig:cn_evolution} an example of the time evolution of defect fraction of each type.  In general, the defect fraction is defined as the (number of defects)$/$(number of particles in the bulk); see Appendix \ref{sec:CountingDefects} for precise methodology of how defect counting was done - it depends on the presence or absence of the boiling layer.

\begin{figure}[h]
\centering
\subfloat[Process 1 - mergers.  In the top panel, we display a double merger - two 5-defects and two 7-defects merge and annihilate each other.   In the bottom panel we display a merger in which there are two pairs of 5- and 7-defects before the merger, and one pair after the merger.]{%
\begin{minipage}{0.7\linewidth}
\centering
\includegraphics[width=\linewidth]{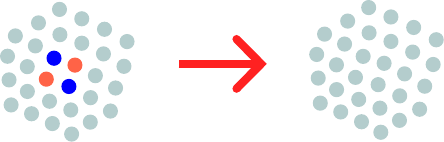}\\[0.5em]
\includegraphics[width=\linewidth]{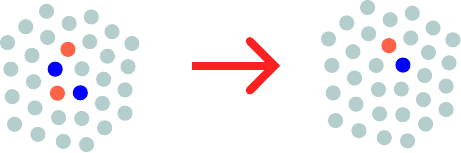}
\end{minipage}
\label{fig:cn_processes1}}
\par
\subfloat[Process 2 - migration.  Here we display two sets of defect pairs migrating.]{%
\includegraphics[width=0.7\linewidth]{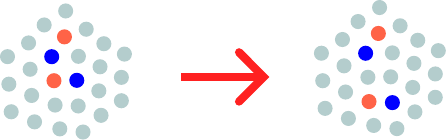}
\label{fig:cn_processes2}}
\par
\subfloat[Process 3 - pair production.  This is the time-reverse of a merger, and involves a production of a pair of 5- and 7-defects.]{%
\includegraphics[width=0.7\linewidth, keepaspectratio]{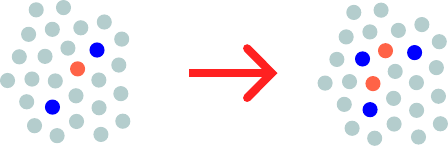}
\label{fig:cn_processes3}}
\caption{Defect processes - mergers, migration, and pair creation.}
\end{figure}

We found that coordination number defects evolve through specific processes: mergers, production, and migration.  Examples of these are displayed in Figs.~\ref{fig:cn_processes1}-\ref{fig:cn_processes3}. 

\section{Defect-free sub-systems} 
\label{sec:annealing}
The aging was discovered in our prior study \cite{PreviousPaper}, when we analyzed velocities and found that there is residual creeping motion left after the breathing transient.  Our study here revealed the mechanism of this aging process - gradual elimination of coordination number defects by means of mergers and migration events.  However, we found that such defects - and their gradual decrease through these processes - is not unique to a delayed system that displays breathing.  In fact, similar dynamics happen for the undelayed system, and even for pure swarmers that have no phase degrees of freedom at all.  This latter system's evolution is just overdamped dynamics in a potential landscape.  We wanted to see if breathing transients that are specific to the delayed swarmalator system causes the system to self-anneal during the breathing stage; whether repeated expansion–contraction cycles give the cluster opportunity to rearrange more effectively during the transient, ultimately lowering the final density of defects in the bulk.  To test this hypothesis, we measured the the final defect fraction as a function of the quality factor (which is effectively a measure of the number of oscillations during the breathing stage).  

\begin{figure}[h]
\centering
\subfloat[Defect fraction vs Q-factor.  One dot per initial condition.  Simulations were done for $20$ initial conditions per each $\tau$, which ranged between $5$ and $20$ in steps of $0.5$.]{%
\includegraphics[width=\linewidth]{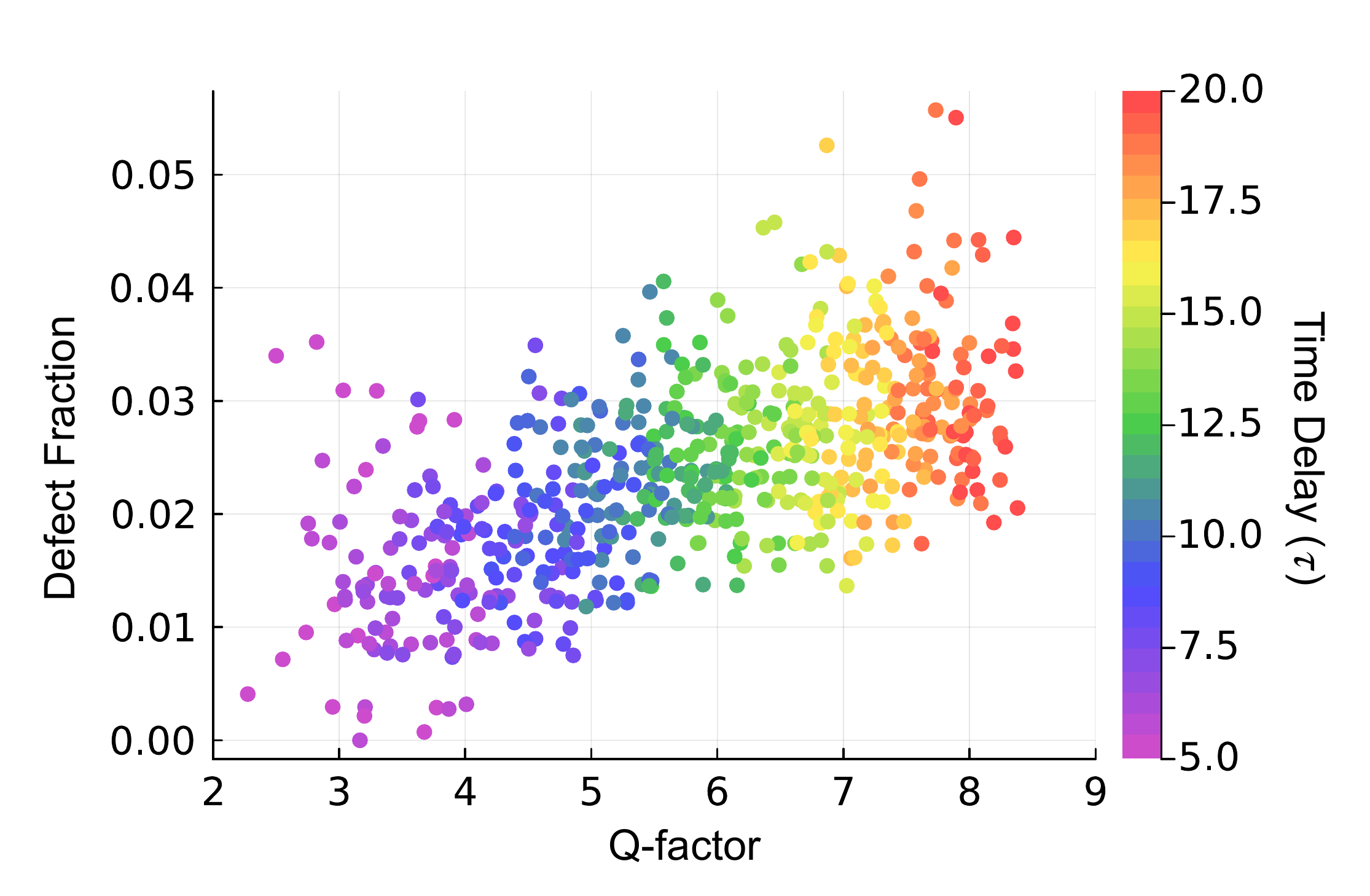} 
\label{fig:df_vs_qf}}
\par
\subfloat[Average and standard deviation computed over bins of size $\Delta Q = 0.5$, from $Q=2.5$ to $Q=9$.]{%
\includegraphics[width=0.9\linewidth]{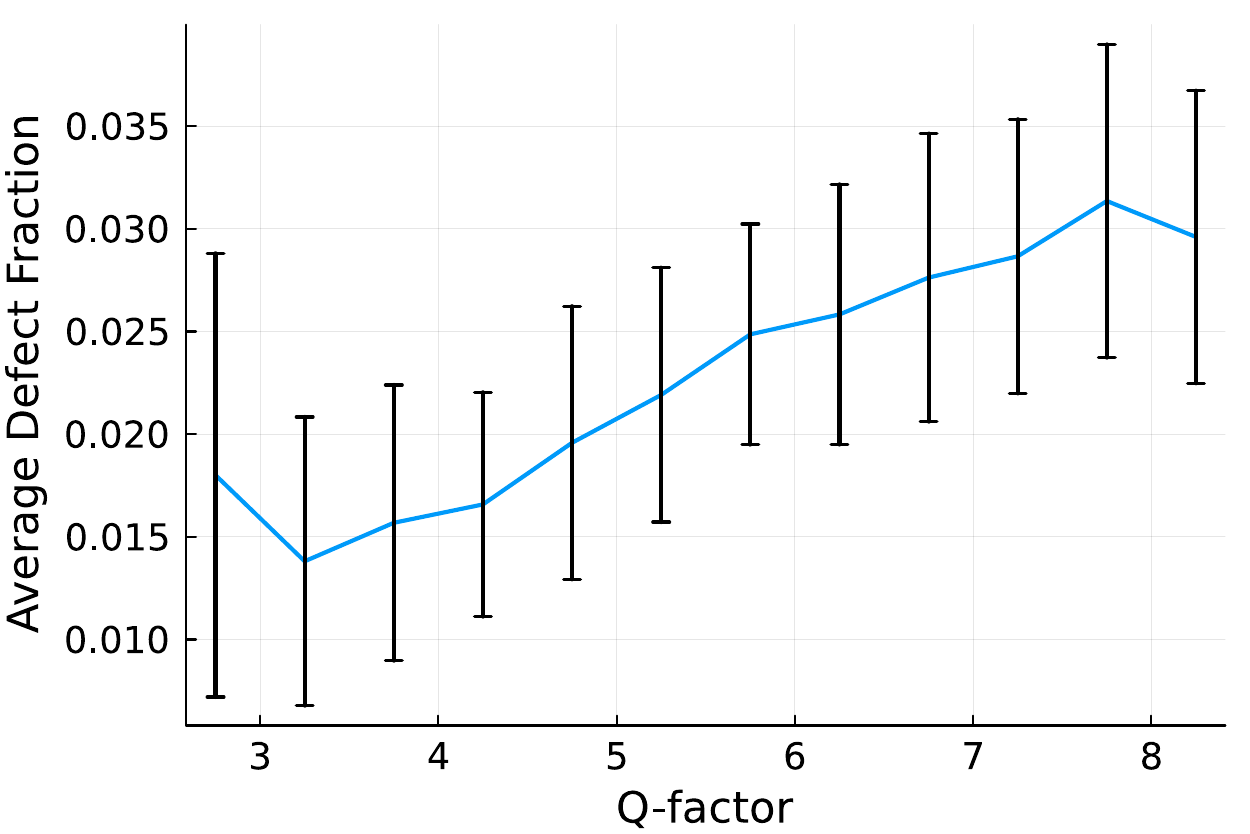} 
\label{fig:avg_df_vs_qf}}
\caption{Dependence of the fraction of defects (computed according to the method described in Appendix \ref{sec:CountingDefects}\ref{sec:FinFraction}) on the quality factor of the breathing during the transient.  The parameters are $N=1000$, $J=1$, $K=-0.75$.  For these parameters, $\tau_c \approx 13$.  Simulation time is $70000$.}
\label{fig:fraction-vs-Q}
\end{figure}

Since $Q$ depends on $\tau$, we performed simulations of system with different delays.  For each $\tau$, we simulated systems with 20 different random initial configurations in both spatial positions of particles and their phases.  Note that a $Q$-value is not always the same for each $\tau$, and the final defect fraction was also not the same, as both depend on the initial configuration (the role of initial conditions was briefly studied in Section 9.1 of \cite{blum2021swarming}).  The result of this investigation is presented in Fig.~\ref{fig:fraction-vs-Q}.  

\begin{figure}[h]
\centering
\includegraphics[width=1\linewidth, keepaspectratio]{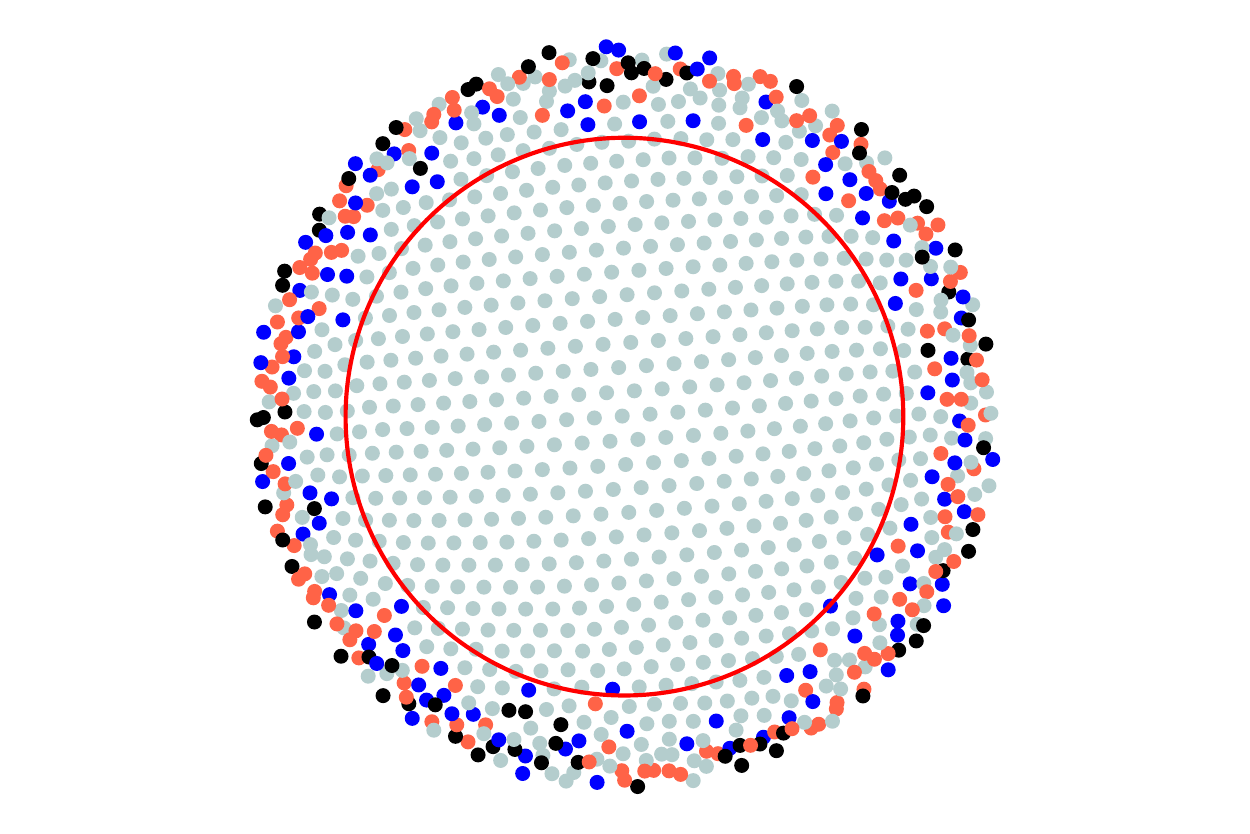}
\caption{Example of a defect-free solid core, where the surface develops a  fluidized boiling layer. Here $N=1000$, $J=1.0$, $K=-0.75$, $\tau=5.5$.  This is the last snapshot of the \href{https://o365coloradoedu-my.sharepoint.com/:v:/g/personal/chku8451_colorado_edu/EcpCa2L3vO5PvAbmbT37RRUBEyADFy4zwyjQtdPAtmyjHA?e=EQKGC0}{Supplementary Video 2}, and corresponds to the system whose defect fraction in Fig.~\ref{fig:fraction-vs-Q} was found to be zero.}
\label{fig:example_defect_free}
\end{figure}

The details of the computational procedure that went into generating these results can be found in Appendix \ref{sec:CountingDefects}.  We are forced to conclude that there does not seem to be an obvious support for the self-annealement hypothesis.  The final fraction of defects shows the trend opposite of what we would expect if the self-annealement hypothesis held true; the fraction of defects decreases with decreasing quality factor. In fact, at lower $Q$-values, we even found completely defect free examples;  these happen for $\tau$ sufficiently below $\tau_c$, in which there is a substantial boiling layer, with a solid core underneath.  Fig.~\ref{fig:example_defect_free} shows an example of such a cluster;  see also the end of the \href{https://o365coloradoedu-my.sharepoint.com/:v:/g/personal/chku8451_colorado_edu/EcpCa2L3vO5PvAbmbT37RRUBEyADFy4zwyjQtdPAtmyjHA?e=EQKGC0}{Supplementary Video 2} for $\tau=5.5$.  This complete expulsion of defects appears to be subject to fluctuations; as the boiling layer perturbs the solid core below, the defects exchange between the two layers (see the \href{https://o365coloradoedu-my.sharepoint.com/:v:/g/personal/chku8451_colorado_edu/EcpCa2L3vO5PvAbmbT37RRUBEyADFy4zwyjQtdPAtmyjHA?e=EQKGC0}{Supplementary Video 2} for $\tau=5.5$). 
The boundary of the solid core is a matter of computational convention (see Appendix \ref{sec:CountingDefects}\ref{sec:sol_core}), and it may be possible for defects to occasionally cross this conventional boundary in both directions.  Therefore, the lower-$\tau$ part of Fig.~\ref{fig:fraction-vs-Q} will somewhat depend on the simulation time.
 
\section{Discussion}
\label{sec:discussion}
In this paper, we have picked up one of the unanswered questions from our prior work \cite{PreviousPaper} - what is the nature of the slow creep that occurs after the breathing has ceased?  We now understand the answer - it is the aging dynamics of the system that takes place through a gradual elimination of coordination number defects.  In fact, we found that this aging happens not only above $\tau_c$, but even in the solid cluster below $\tau_c$.  It is a process that unfolds on timescales much greater than the breathing period or decay timescale of the transient. 

Annihilation occurs between defect pairs, specifically between a 7-defect and a 5-defect.  The same can be said for creation - new defects are born in pairs of a 5-defect and a 7-defect.  If there is a group consisting of three 7-defects and two 5-defects, the total defect excess is $+1$ - and it remains $+1$ if there are annihilation or creation events in that group, based on our observations.  Therefore, the defect excess of a group of defects is a conserved quantity.  Based on the existence of such a conserved quantity, as well as the structure of the phase of the local hexatic order parameter around defects (which tend to occur in groups), these coordination number defects appear to be topological. 

Our study has raised deeper questions that remain unanswered.  First, what is the effective interaction between the defects?  Second, what triggers annihilation and creation events?  For example, does creation of a pair of defects (5 and 7) require an existing defect in the neighborhood?  This appears to be the case, but a clear answer requires a statistical study of many defect creation events.  If a nearby defect is indeed required, is there an effective strain-like quantity that causes the creation of a pair of defects?  The \href{https://o365coloradoedu-my.sharepoint.com/:v:/g/personal/chku8451_colorado_edu/ETMGSMHbAf5ImlFJBXgitzUB9tE3rrig59jxCiwcMLCBZw?e=ucA99i}{Supplementary Video 1} for $\tau=20$ makes it clear that a good fraction of defects disappear not through annihilation events, but via migration to the boundary in pairs.   What is the effective force that pushes defect pairs to the edge?  When there is a thick boiling layer for $\tau$ sufficiently below $\tau_c$, it is possible to catch a system with no defects such as in Fig.~\ref{fig:example_defect_free}.  A tantalizing research question is whether the existence of a ``soft'' outer layer that facilitates the removal of defects from the solid core presents a unique strategy for perfect crystallization in other active matter systems; is it possible to engineer active particle interactions which cause the formation of a layer that absorbs defects, while leaving other parts of the system defect-free?  Going back to the swarmalator system specifically, it is also an interesting question whether it is in principle possible to end up with a defect-free system for $\tau > \tau_c$ - perhaps for some very special initial condition?  Finally, it's also worth asking what might happen on much longer time scales and for bigger systems.  All these questions emerged as we delved into the problem posed in the previous paper  \cite{PreviousPaper}.

We now put our work in a broader perspective.  We provided an example of a two-dimensional non-equilibrium system that ages by means of a gradual elimination of topological defects via mutual annihilation and migration to the boundaries.  We mentioned that aging by elimination of defects happens also in a non-delayed model, and even in a basic swarm.  Aging via mutual defect annihilation is known to occur in other systems as well.  Coarsening processes that involve mutual annihilation of topological defects - typically vortices and antivortices - have been reported in nematic liquid crystals \cite{chuang1993coarsening}, the XY model \cite{yurke1993coarsening}, ferromagnetic superfluids \cite{williamson2019anomalous}, active fluids made from Quincke rollers \cite{chardac2021emergence}, two-dimensional complex Ginzburg-Landau equation \cite{liu2020aging, arnold2022dynamics}, and even in a system of Kuramoto oscillators \cite{rouzaire2022dynamics}.  A system that shares a significant overlap with the phenomenology of our model is described in the recent work by Digregorio and co-authors \cite{Cugliandolo_2Dstuff}, which proposed a unified analysis of passive and active repulsive Brownian disk systems in two dimensions.  Their framework is specific to power-law interactions - which required a box to confine the particles.  Our system has all-to-all attractions, and power-law repulsions;  therefore, it forms a cluster that does not require a box.  The class of systems considered in \cite{Cugliandolo_2Dstuff} is based on potential with or without an additional active force; our system is non-Hamiltonian. 

We pause here to reflect on the significance of boundaries.  In addition to defect-defect annihilation - which does take place in all of the aforementioned sources during the aging process - our system also allows expulsion of defects into boundaries.  We saw that for a delay below $\tau_c$, the boundary develops a fluidized boiling layer - and a sufficiently thick boiling layer allows a complete absorption of defects, leaving the solid core defect-free.  This finding appears quite unique to soft-matter and dynamical systems, although reminiscent processes occur in irradiated materials at high temperature \cite{hung2022insight, aitkaliyeva2017irradiation, zheng2022grain, doan2003elimination}, where grain boundaries, surfaces, and dislocation cores can serve as sinks for defects (see also a different system here \cite{gerbode2010glassy}).  However, the similarity here may be superficial, due to these systems being rather different from ours.  The complete elimination of defects resulting from the interplay of the free boundary and delayed dynamics calls for further search of similar behavior in other models.  If this phenomenon could be generalized to other systems - perhaps also requiring delayed dynamics - it could pave way for a new mechanism of constructing defect-free materials and systems of active particles.  

\begin{acknowledgements}
The authors gratefully acknowledge useful discussions on defect dynamics with David R. Nelson and Leticia Cugliandolo.  TC acknowledges the funding support from the NSRF via the Program Management Unit for Human Resources \& Institutional Development, Research and Innovation [grant number B39G680007], and the computational resources from Chula Intelligent and Complex Systems Lab. CK acknowledges the funding support in part by the Interdisciplinary Quantitative Biology (IQ Biology) PhD program at the BioFrontiers Institute, University of Colorado Boulder, and the National Science Foundation NRT Integrated Data Science Fellowship [award 2022138].  OK acknowledges startup funds provided by Queens College, City University of New York.  The authors used OpenAI ChatGPT for assistance with wording, structural editing, and conceptual clarity; all substantive content is the authors’ own.
\end{acknowledgements}

\appendix
\section{Numerical simulation details}
\label{sec:SimDetails}
All of the simulations in this article have been done using Julia programming, with the exception of Fig.~\ref{fig:Fig1}, which was done using \textbf{dde23} in Matlab. In Julia, we mainly use the differential equation solver package, \emph{DifferentialEquations.jl}, to solve delayed differential equations. The algorithm that we used to solve the differential equation is \textbf{Tsit5} or Runge-Kutta Pairs of Orders 5(4) \cite{Tsitouras2009-hw}. In addition, fixed the random seed while exploring the effect of varying a given parameter. This process initialized the system at the same initial conditions and allowed us to observe the effect of varying each parameter on the system.   The exception concerns results in Fig.~\ref{fig:fraction-vs-Q}, where $20$ different random seeds ($20$ different initial conditions) were used per $\tau$.

Github link for our code is provided in \cite{DelaySwarmJulia}.

\section{Tesselation procedure}
\label{sec:Tesselation}

To evaluate the hexatic order parameter $\Psi_j$ of particle $j$, we first identify its nearest neighbors using the Delaunay triangulation algorithm \cite{DelauneyRef}. This approach ensures a well-defined local connectivity based on spatial proximity. Next, we compute the angles between the horizontal axis and the vectors connecting particle $j$ to each of its neighbors $k$, using the two-argument arctangent function \emph{atan2}, which is natively available in Julia programming. These angles are then used in the computation of the order parameter via Eq.~(\ref{eq:Psi_def}).

\section{Process to count the number of defects}
\label{sec:CountingDefects}

The method of counting defects differs depending on whether the system is in the boiling ($\tau < \tau_c$) or in the quasistatic ($\tau > \tau_c$) regime.  Being in the quasistatic phase allows for a direct evaluation of the bulk radius and defect count within it.  Conversely, boiling regime is characterized by persistent (mostly radial) motion of particles in the outer annulus of the system, with a solid core below.  Defects only have a meaning in the solid core, so we must estimate its radius. 

\subsection{Bulk radius}
\begin{figure}
\centering
\includegraphics[width=0.45\textwidth,keepaspectratio]{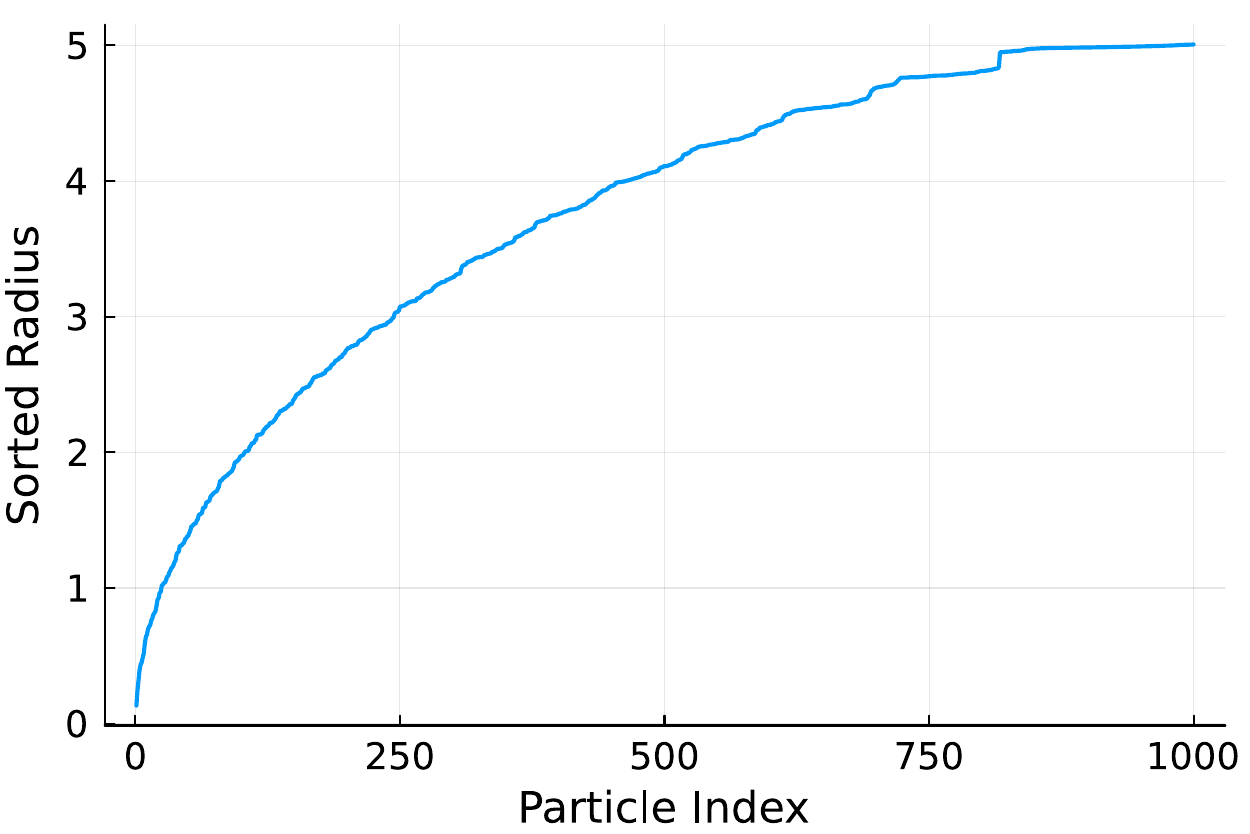}
\caption{Sorted particle radii for $N=1000$, $K=-0.75$, $J=1.0$, $\tau = 20$.}
\label{fig:sorted_r}
\end{figure}

This procedure was used for $\tau > \tau_c$.  Fig.~\ref{fig:sorted_r} is the plot of the radius of each particle in a $1000$-particle system.  The reason for the specific form of this plot is the following.  We showed in \cite{PreviousPaper} that the density (number of particles per area) follows closely the relationship $(1-r/R)^{-1/2}$.  Thus, close to the center, the density is approximately constant \cite{PreviousPaper}, so the number of particles $n$ within a radius should scale like $r^2$ or $r \propto \sqrt{n}$.  Closer to the edge of the cluster (at radius $R$), the density rapidly grows, so the plot should become horizontal.  In addition, due to the discrete nature of particles and their interactions, they organize into ring structures, which is especially prominent near the edge (see, for instance, Fig.~3 in \cite{PreviousPaper}).  For this reason, the radius versus the particle number displays step-like features for $r$ close to $R$, with the biggest jump at the last ring.  This was used to identify the bulk.  We computed the slope of radius vs.~particle index function, and found the radius at which the slope is greatest - this is the transition to the last step - the outer ring, or the ``crust'' of the cluster.  We used $85\%$ of this value as the criterion for the bulk radius.
This particular process has been applied the same way in every setup with a quasistatic state.

\subsection{Solid core}
\label{sec:sol_core}

\begin{figure}
\centering
\subfloat[Standard deviation of $\Delta r$ over a time series, as a function of particle index.  In this figure, we use $N=1000$, $J=1.0$, $K=-0.75$, $\tau = 5.0$. The red vertical line is the location of the calculated solid core radius (while the reported value is $85\%$ of it).]{%
\includegraphics[width=\linewidth,keepaspectratio]{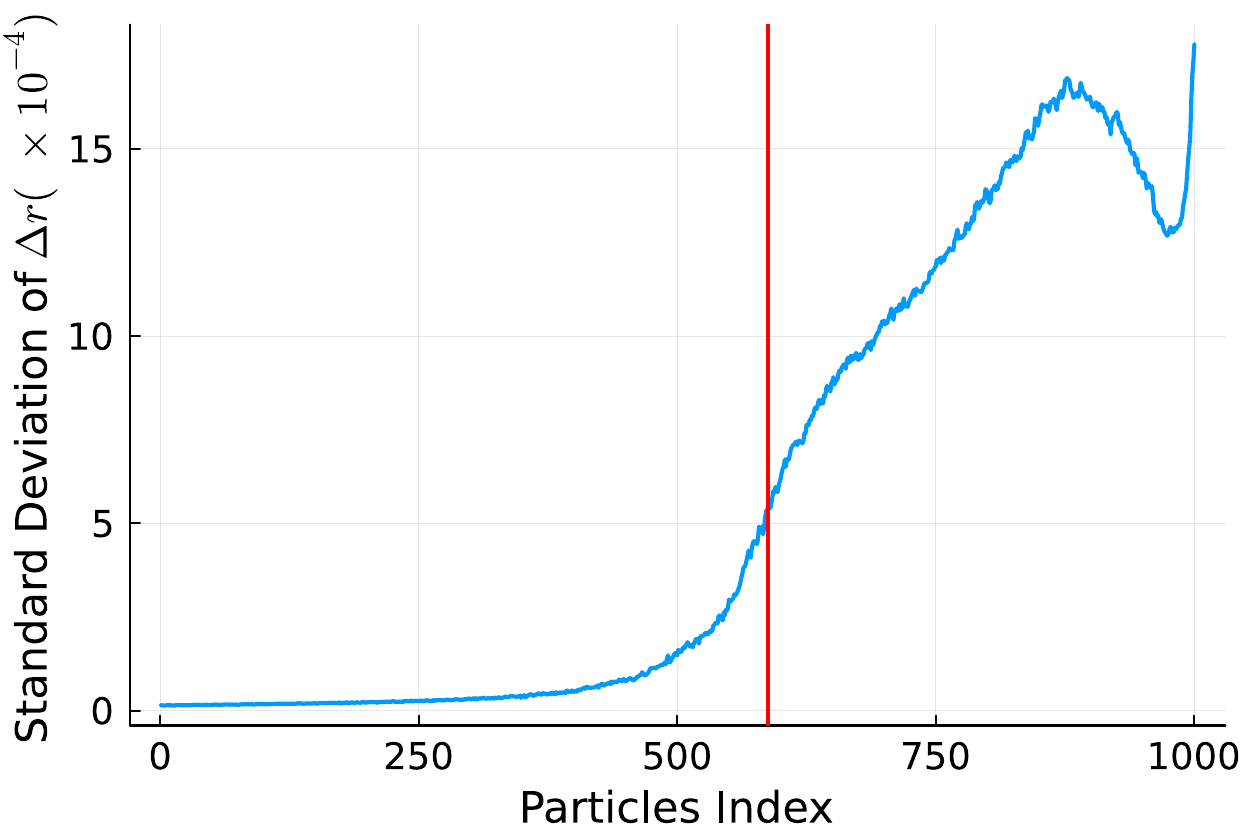}
}
\par
\subfloat[Six point estimate of the slope of $std(\Delta r)$ vs. particle index.  The red vertical line is the location of the calculated solid core radius (while the reported value is $85\%$ of it).]{%
\includegraphics[width=\linewidth,keepaspectratio]{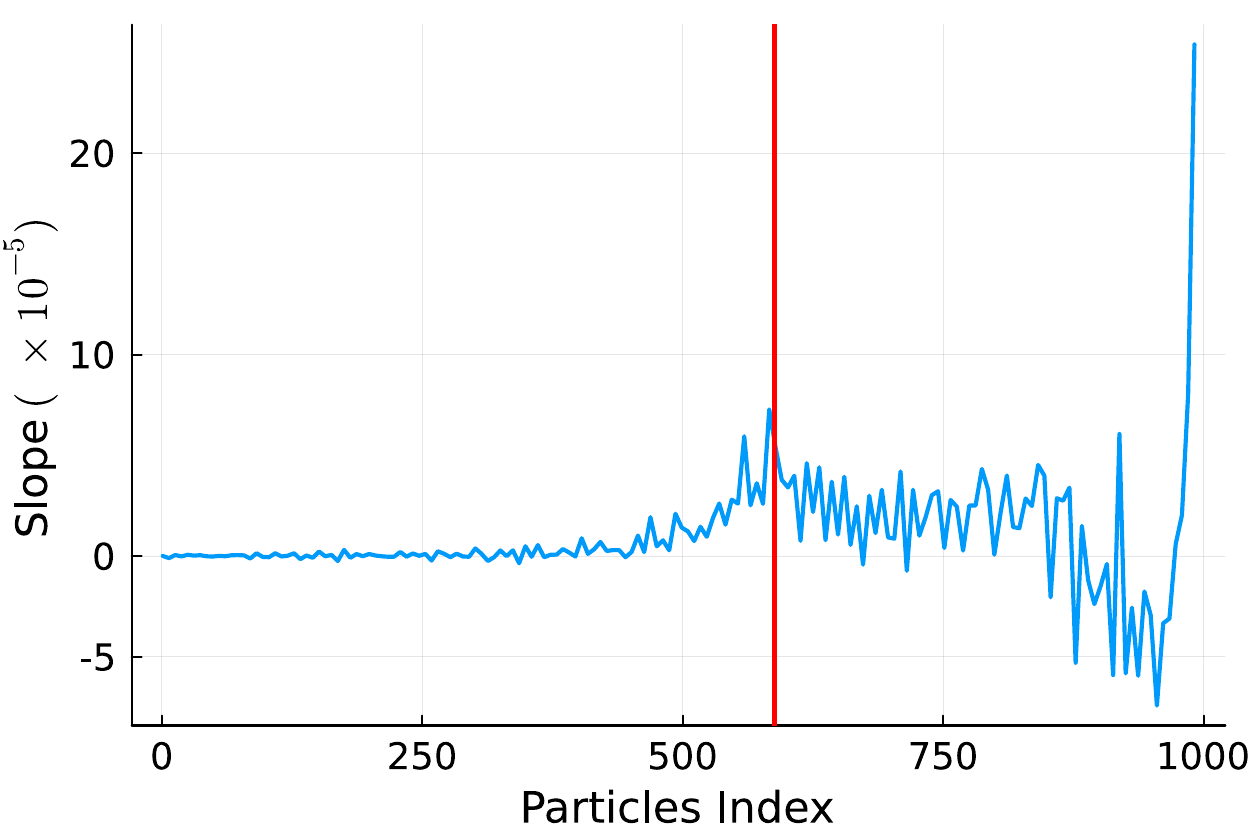}
}
\caption{Calculation of the solid core radius - an example.}
\label{fig:diff_std}
\end{figure}

When the system is in the boiling state for $\tau < \tau_c$ we need to identify the solid core, which lies below the fluidized layer.  To do this, we implemented the following algorithm at the last 5000 time steps of the simulation: (i) at each time step, we sort particle radii, (ii) we measure the difference between two consecutive time steps, $\Delta r$, (iii) we obtain a time series of $\Delta r$ over the last 4999 time steps of the simulation and compute the standard deviation over this time series as a function of the particle index; an example is displayed in Fig.~\ref{fig:diff_std}.  We are interested in the index at which this function has the largest slope.  Measuring local slopes, however, tends to amplify small fluctuations and did not always give an accurate prediction of the solid core radius.  Thus, (iv) we measured the slope over six points (effectively, an average over six neighboring local slopes), and sought the location at which this locally-averaged slope attains a maximum (and related it to radius).  

\begin{figure}[h]
\centering
\includegraphics[width=1\linewidth]{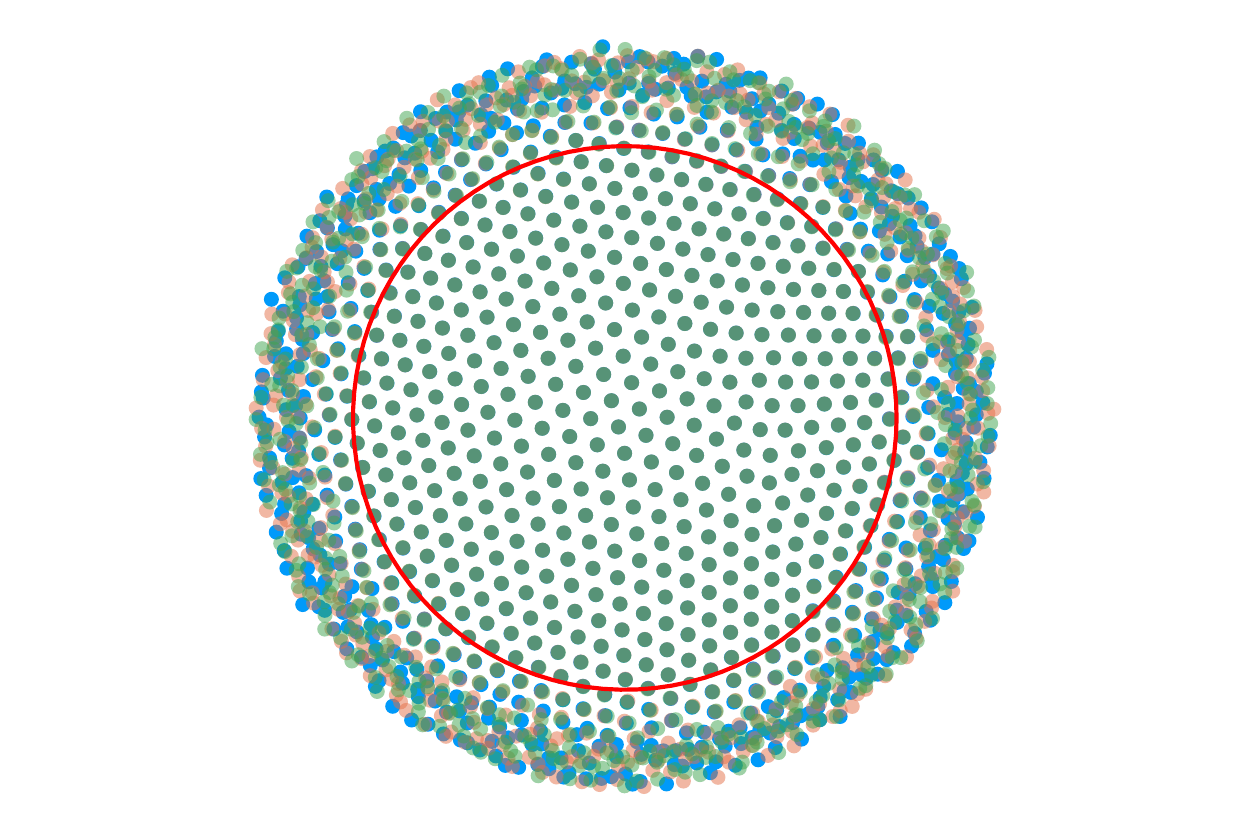}
\caption{Superposition of multiple snapshots (in different colors) in the boiling state, showing the  fluctuations in the boiling layer, and stable structure that forms the solid core. The red circle represents the estimated bulk radius, which is $85\%$ of the radius corresponding to the red line in Fig.~\ref{fig:diff_std}; it is the result of the procedure outline in Appendix  \ref{sec:CountingDefects}\ref{sec:sol_core}.  Here the parameters are $N=1000$, $J=1.0$, $K=-0.75$, and $\tau=5.0$.}
\label{fig:rbulk_solid_core_ex}
\end{figure}

As in the case of the bulk radius estimation, we report $85\%$  of the value obtained with the described algorithm.  From our experience, this whole procedure gave an accurate assessment of the radius of the solid core.  An example is given in Fig.~\ref{fig:rbulk_solid_core_ex}.  We found that a step in the radius vs.~the particle index that we use to identify the bulk radius for $\tau>\tau_c$ actually persists for $\tau$ somewhat below $\tau_c$.   
For $(J, K)$  parameters that were used to construct Fig.~\ref{fig:fraction-vs-Q} (when $\tau_c \approx 13$), this step persists down to $\tau \approx 9$.  We found this by checking that using the solid core procedure between these $\tau$ values produces essentially the same boundary.  So, we used the bulk radius procedure above this $\tau$ and a solid core radius estimation below it.  This gave us accurate estimations of the solid core, with a faster computation.

\subsection{Calculating the defect fraction}
\label{sec:FinFraction} 
\begin{figure}[h]
\centering
\includegraphics[width=1.0\linewidth]{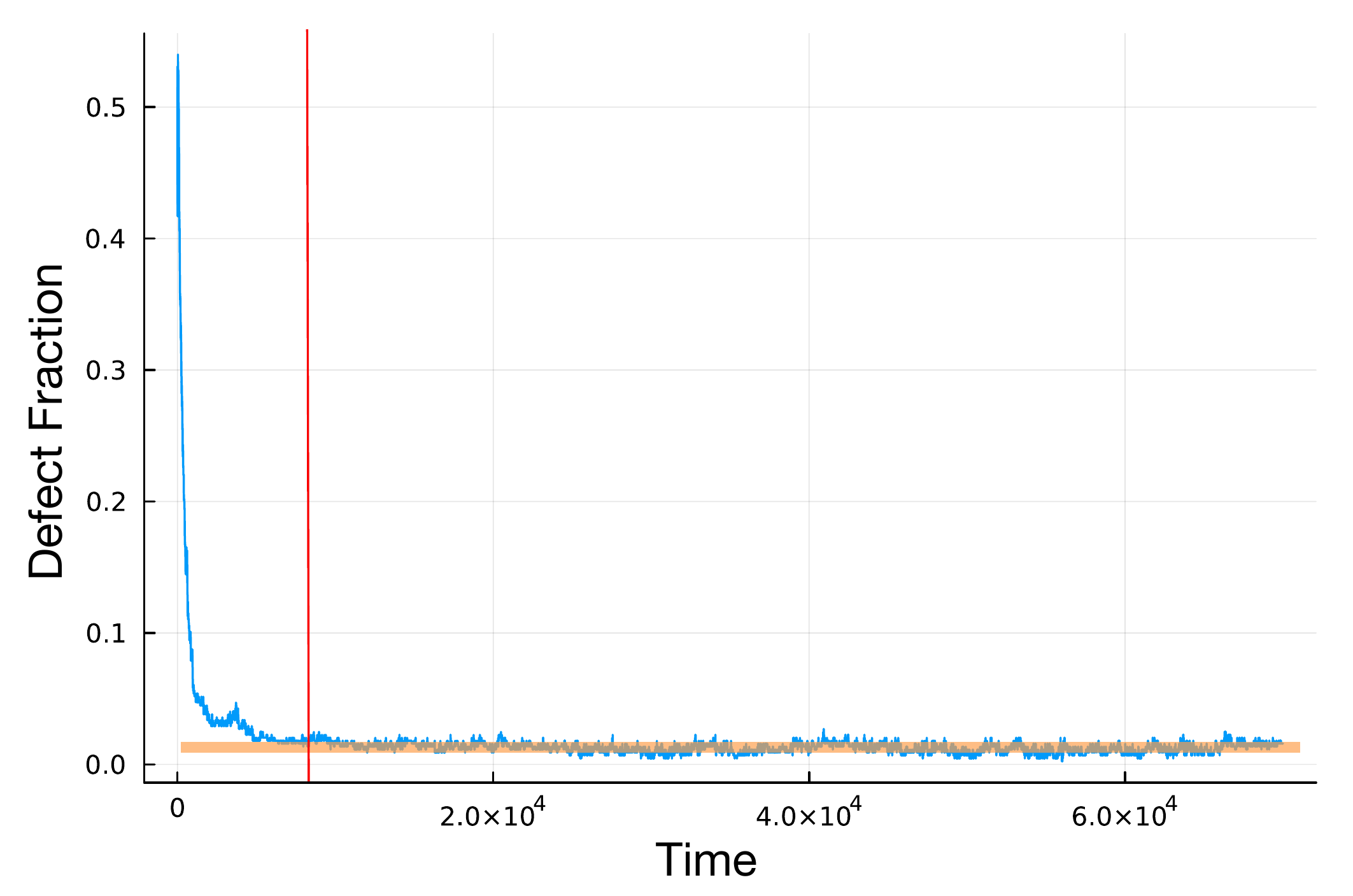}
\caption{Defect fraction over time, an example.  The horizontal yellow band denotes the range of fluctuations of the defect fraction within one standard deviation at late times.  The defect fraction as a function of time first enters this band at the time denoted by the vertical red line.  The reported final defect fraction is average over the subsequent $500$ time steps.  In this example, $N=1000$, $=1.0$, $K=-0.75$, and $\tau = 5.5$.} 
\label{fig:final_df_process}
\end{figure}

The defect fraction decreases over time before stabilizing at a plateau above $\tau_c$.  In the boiling regime (below $\tau_c)$, there are residual fluctuations (see \href{https://o365coloradoedu-my.sharepoint.com/:v:/g/personal/chku8451_colorado_edu/EcpCa2L3vO5PvAbmbT37RRUBEyADFy4zwyjQtdPAtmyjHA?e=EQKGC0}{Supplementary Video 2} for $\tau=5.5$). To estimate the final defect fraction, we use the following protocol.  We compute the average and standard deviation of the defect fraction over a sufficiently long interval near the end of the simulation well beyond the onset of the aging phase.  Based on these statistics, we define a tolerance band (visualized as a horizontal colored band in Fig.~\ref{fig:final_df_process}) and locate the earliest time at which the defect fraction enters this band (vertical red line in Fig.~\ref{fig:final_df_process}). From this point onward, we average the defect fraction over the subsequent $500$ time steps to reduce statistical fluctuations and obtain an estimate of the final defect fraction value.
\section{Damped averaged radius oscillation frequency and decay rate}
To estimate the quality factor of the breathing oscillation, we fit the radius vs.~time with a functional form for a damped harmonic oscillator 
\begin{align}
y(x) &= Ae^{-\gamma x}\sin(\omega x +\phi) \label{eq:damped_osc},
\end{align}
Initially, the algorithm fits the entire dataset using Eq.~(\ref{eq:damped_osc}). It then computes the mean squared error (MSE) between the fitted curve and the empirical data. If the MSE exceeds a predefined threshold, the algorithm iteratively shortens the fitting interval from the beginning of the time series. This strategy is motivated by the observation that the swarmalator system exhibits chaotic behavior before transitioning into a coherent breathing phase. The iterative trimming continues until the breathing state is isolated and a satisfactory fit is achieved.  Once optimal fitting parameters are obtained, the quality factor $Q$, which characterizes the number of oscillations during one decay time is calculated as $Q = \omega/(2\gamma)$.

\bibliography{main_final.bib}

\end{document}